\documentclass[aps,pre,showpacs,preprint,floatfix]{revtex4}%
\usepackage[dvips]{graphicx}
\usepackage{amsmath}
\usepackage{amsfonts} 
\usepackage{amssymb}%

\DeclareGraphicsExtensions{.eps}

\begin{document}
\title{Generalization of the reaction-diffusion, the Swift-Hohenberg and the Kuramoto-Sivashinsky equation and effects of finite propagation speeds}
\date{\today}
\author{Axel Hutt}
\email{ahutt@uottawa.ca}
\affiliation{Department of Physics, University of Ottawa,150 Louis Pasteur, Ottawa, Ontario, K1N-6N5, Canada}

\begin{abstract}
The work proposes and studies a model for one-dimensional spatially extended systems, which involve nonlocal interactions and finite propagation speed.
It shows that the general reaction-diffusion equation, the Swift-Hohenberg equation and the general Kuramoto-Sivashinsky equation represent special cases of the proposed model for limited spatial interaction ranges and for infinite propagation speeds. Moreover, the Swift-Hohenberg equation
is derived from a general energy functional. After a detailed validity study on the generalization conditions, the 
three equations are extended to involve finite propagation speeds. Moreover linear stability studies of the
extended equations reveal critical propagation speeds and novel types of instabilities in all three equations.
In addition, an extended diffusion equation is derived and studied in some detail with respect to finite propagation speeds. The extended model allows for the explanation of recent experimental results on  
non-Fourier heat conduction in non-homogeneous material.
\end{abstract}
\maketitle

\section{Introduction}
The propagation of activity in spatially extended systems has attracted much attention in the last centuries, starting from celestial mechanics in the sixteenth century to the study in complex physical, chemical or biological systems~\cite{Haken04book,Coombes05_2} in the last decades. In such systems the communication between the system subunits plays a 
decisive role. These subunits may be coupled locally to their next neighbors or may reveal longer-distance connections. In the latter
case, the finite speed of interactions, i.e. the finite propagation speed, may yield temporal delays which affect the space-time dynamics of the system. Such effects have been found in several systems, e.g. in neural networks~\cite{HakenBook02,Eurich02,Laing+Coombes06,Hutt+Frank05}, quantum devices~\cite{Martin+Landauer92}, porous and non-homogeneous media~\cite{Tzou+Chen98,Mitra_etal95},
in metals irradiated by laser pulses~\cite{Klossika_etal96}, hot plasma~\cite{Lazzaro+Wilhelmsson98} and electronic circuits~\cite{Weller97,Pountourakis03} as the internet or computational networks. \\
In order to describe spatial systems mathematically, partial differential equations (PDE) have been applied 
widely. However, most PDE models neglect effects caused by finite propagation speeds in the system. To consider
 these effects in PDE models, various approaches have been followed as, e.g. the introduction of additional 
temporal derivatives~\cite{Cattaneo58} or temporal constant delays~\cite{Glass+Mackey_Book88}. 
However, propagation delays depend on the distance between two locations and thus represent
space-dependent delays. In contrast to the PDE formulation, a natural way to consider these delays are
 integral-differential equations (IDE) which sum up all activity in a spatial domain and easily consider
 propagation delay, see e.g.~\cite{Gerstner95,Crook97}. Moreover, the strong connection between PDE and 
 IDE models is well-known~\cite{Murray}. The present work proposes a novel IDE model involving finite  
propagation delays which generalizes well-studied PDE models in one spatial dimension, namely the
 reaction-diffusion models, the Swift-Hohenberg model and the Kuramoto-Sivashinsky model.
 Moreover re-calling the mathematical description of neuronal populations by IDEs, the proposed 
model represents a generic pattern forming model with a vast range of possible applications.\\
In addition to the generalization described above, the novel IDE model allows for the extension of PDE models 
in order to consider propagation delay effects. The present work shows that the 
results obtained in IDE models involving finite propagation speeds can be applied easily to PDE models. Hence 
this work extends previous studies~\cite{Hutt+Atay_PhysicaD05,Atay+Hutt_SIAM05} as it clearly distinguishes 
local dynamics from nonlocal spatial interactions and elicits the strong connection between well-known PDE-models of pattern forming systems and the IDE models in neural field
 theory~\cite{Ermentrout98,Jirsa04_Rev,Bressloff97}. We derive novel extensions of the diffusion equation,
 the Swift-Hohenberg equation (SHE) and the Kuramoto-Sivashinsky equation now involving finite propagation speeds.
These extensions are important if the speed of the propagating activity in the system approaches
the propagation speed of the system. Such ultra-fast phenomena have been observed experimentally 
in solids~\cite{Fujimoto_etal99,Rieger04}, in plasma~\cite{Lazzaro+Wilhelmsson98} and on solid and fluid surfaces~\cite{Bovensiepen06,Aliev95}.\\
 The present work is structured as follows. Section~\ref{sec_general} introduces the integral-differential 
 equation model and elicits its formal relation to partial differential equations. Then the subsequent section 
shows the reduction procedure to the PDE models and investigates the validity of that
 reduction. Eventually section~\ref{sec_stability} studies the linear behavior of three specific models and 
examines their dependence on the finite propagation speed, while the last section closes the work
 by a brief discussion of the obtained results. 
 
\section{The general model}~\label{sec_general}
The present work considers the general evolution equation
\begin{eqnarray}
&&\hat{T}V(x,t)=h[V(x,t)]\nonumber\\
&&+\int_{\Omega} dy\,K(x-y)f[V(y,t-\tau(x-y))]\nonumber\\
&&+\int_{\Omega} dy\,L(x-y)g[V(y,t-\tau(x-y))]\label{eqn_final}
\end{eqnarray}
with the scalar field variable $V(x,t)$ and the spatial domain $\Omega$ which is assumed being 
large but finite. The temporal operator $\hat{T}=\hat{T}(\partial/\partial t)$
represents the temporal linear dynamics of an uncoupled element at spatial location $x$. For instance,
a damped oscillator would be modeled as $\hat{T}V=\partial^2 V/\partial t^2+\partial V/\partial t+V$.
In the following, we specify the operator by $\hat{T}(0)=1$. In addition, the self-interaction 
$h[V(x,t)]$ may 
represent an additional nonlinear driving. Considering the previous example, the nonlinear damped
oscillators may be modeled by $\hat{T}V=V-\sin(V)$, i.e. $h[V]=V-\sin(V)$. Moreover, to extent the system
by spatial coupling, $K(x-y)$ and $L(x-y)$ in Eq.~(\ref{eqn_final}) represent the different coupling
functions between elements at spatial locations $x$ and $y$ and the two coupling functions
belong to two different spatial interaction types. For instance, in neural nets neurons may be
coupled by excitation and by inhibition, which is modeled typically by an excitatory net and an inhibitory 
net with $K(x-y)>0$ and $L(x-y)<0$, respectively.  In the next section, this distinction turns 
out out to be very important in the context of the Kuramoto-Sivashinsky equation. In addition, the
 functionals $f[V]$ and $g[V]$ allow for various (nonlinear) interaction types of the corresponding
 spatial interactions. Finally the propagation delay $\tau(x-y)=|x-y|/c$ with the propagation speed $c$ 
takes into account the finite time it takes the signal to propagate from one spatial location $x$ to
 another location $y$.\\
We point out that Eq.~(\ref{eqn_final}) is similar to previous IDE models of neuronal populations~\cite{Ermentrout80,Hutt_Network03}, which can be derived from basic 
properties in neural tissue~\cite{Hutt+Atay_PhysicaD05}. In addition, it extends the previous IDE 
models by considering local interactions and an additional spatial interaction type and thus 
renders the model more realistic for physical systems. As we show in the subsequent section, this
 extension proves to be very powerful.

Finally recall the relation of IDEs to PDEs. Some previous studies remarked that IDEs generalize PDEs~\cite{Murray,Hutt+Atay_PhysicaD05} by the identity 
\begin{eqnarray}
\int_{-\infty}^{\infty}dyK(x-y)S[V(y)]=\sum_{n=0}^\infty(-1)^nK_n\frac{\partial^n S[V(x)]}{\partial x^n}\label{eqn_PDEexp}
\end{eqnarray}
with a nonlinear functional $S$ and the kernel moments $K_n=\int d\eta K(\eta)\eta^n/n!~\forall n\in\mathbf{N}_0$. 
The expansion (\ref{eqn_PDEexp}) represents an order expansion of spatial interactions 
whose order $n$ indicates the spatial interaction range. Hence $K_n$ represent the contribution of 
spatial interactions of order $n$. Typically local or short-range PDE models involve spatial derivatives of 
up to second order, while PDE models with spatial derivatives of up to fourth order involve non-local or
long-range interactions. The expansion (\ref{eqn_PDEexp}) extends this classification to spatial derivatives of arbitrary order.

\section{Reduction to specific models}
This section shows that Eq.~(\ref{eqn_final}) generalizes reaction-diffusion systems, the SHE and the Kuramoto-Sivashinsky equation in one-dimensional spatial systems. Since these equations 
assume an infinite propagation speed in the system, we set $\tau(x-y)=0$. 

\subsection{Reaction-diffusion equations}\label{subsec_RD}
First let us focus on reaction-diffusion models, which allow for the mathematical description of 
diverse spatio-temporal phenomena. We mention the heat propagation or phase transitions in solids~\cite{LL_STAT1,Walgraef96}, chemical reactions in fluids~\cite{Nicolis+Prigogine77,Cross93} and 
the pattern formation in biological  systems~\cite{Haken04book,Murray}. Reaction-diffusion 
equations are obtained from (\ref{eqn_final}) by choosing $\hat{T}=\partial/\partial t$, $g[V]=0$ and $f[V]=f_1V$. Further,
the kernel $K(x)$ is symmetric, i.e. $K_{2n+1}=0$, and exhibits short-ranged spatial interactions with
$K_{n}\to 0$ for $n>2$. Then the application of Eq.~(\ref{eqn_PDEexp}) yields
\begin{eqnarray}
\frac{\partial V(x,t)}{\partial t}=\bar{h}[V(x,t)]+D\frac{\partial^2}{\partial x^2}V(x,t)\label{eqn_RD}
\end{eqnarray}
with $\bar{h}(V)=h[V]+K_0f_1V$ and $D=K_2f_1$. This reaction-diffusion equation~\cite{Cross93} accounts for 
(non-)linear self-interactions represented by $\bar{h}[V]$ and considers local diffusive interactions with
 diffusion constant $D$.
Hence, reaction-diffusion systems neglect spatial interactions of order $n\ge 4$.

\subsection{Swift-Hohenberg equation}\label{subsec_SH}
In their famous work on fluctuations near the onset of the Rayleigh-B\'enard convection~\cite{Swift+Hohenberg,Swift+Hohenberg2}, Swift and Hohenberg derived 
an order parameter equation for the temperature and fluid velocity dynamics of the convection.
This work has attracted much attention in the last decades and the order parameter 
equation has been interpreted as a model system for pattern formation~\cite{Cross93}.
For instance, Lega et al. modeled the spatio-temporal pattern formation in large aspect ratio 
lasers~\cite{Lega94,Longhi+Geraci96} by the SHE and Brazovskii~\cite{Brazovskii75} discussed 
the SHE in the context of the condensation of liquid crystals fluids. However, in spite of its 
broad applicability, the SHE fails to model some experimental 
results~\cite{Cross93} and several previous studies extended the equation by adding 
some terms~\cite{Manneville83,Bestehorn+Haken90}. Hence the application of the 
SHE is not limited to the original physical problem and thus represents 
a generic model for the spatio-temporal dynamics of spatially extended 
systems~\cite{Bestehorn+Haken90}.\\
To gain the SHE from Eq.~(\ref{eqn_final}), we choose the temporal operator to $\hat{T}=\partial/\partial t$ and define $h[V]=aV-bV^3$, $g[V]=0$, $f[V]=f_1V,~f_1<0$. Subsequently
we obtain
\begin{eqnarray}
 \frac{\partial V(x,t)}{\partial t}=h[V(x,t)]+f_1\int_\Omega dy\,K(x-y)V(y,t)~.\label{eqn_SH_nonlocal}
\end{eqnarray}
In addition the kernel $K(x)$ is symmetric 
and exhibits longer-ranged spatial interactions, i.e. $K_{n}\to 0$ for $n>4$. The subsequent application
of Eq.~(\ref{eqn_PDEexp}) yields
\begin{eqnarray*}
\frac{\partial V(x,t)}{\partial t}&=&aV(x,t)-bV^3(x,t)+K_0f_1V(x,t)\nonumber\\
&+&K_2f_1\frac{\partial^2}{\partial x^2}V(x,t)+K_4f_1\frac{\partial^4}{\partial x^4}V(x,t)\label{eqn_sh1}
\end{eqnarray*}
By choosing $b=K_2^2|f_1|/4K_4>0$ and re-scaling time and space according to 
$t\to (4K_4/K_2^2|f_1|)t$ and $x\to\sqrt{2K_4/K_2} x$ respectively, we find the SHE~\cite{Cross93}
\begin{eqnarray}
\frac{\partial V(x,t)}{\partial t}&=&\varepsilon V(x,t)-V^3(x,t)-\left(1+\frac{\partial^2}{\partial x^2}\right)^2V(x,t)\nonumber\\&\label{eqn_SH}
\end{eqnarray}
with $\varepsilon=4K_4a/K_2^2|f_1|-4K_4K_0/K_2^2+1$. Hence, systems obeying the SHE neglect spatial interactions of order $n\ge 6$. \\
Moreover, Eq.~({\ref{eqn_SH}}) maybe derived from an energy functional ${\mathcal F}[V]$ by
the generalized Ginzburg-Landau equation
\begin{eqnarray}
\frac{\partial V}{\partial t}=-\frac{\delta{\mathcal F[V]}}{\delta V} ,\label{eqn_GL}
\end{eqnarray}
Since the present work aims to introduce nonlocal interactions to physical models in a generalized way, 
we follow the approach of Ginzburg and Landau and attempt to model the
 spatio-temporal dynamics of spatially extended systems by a generalized energy functional. 
Taking into account the spatial interactions of elements in a system, the interaction energy in a 
spatial field between two spatial locations $x$ and $y$ at time $t$ reads $K(x-y)V(x,t)V(y,t)$. Here 
the term $K(x-y)$ quantifies the interaction strength between both
elements~\cite{LL_Field}. Considering local interactions $h[V]$ and their corresponding energy 
contributions $W[V]$ additionally, then the field energy is the space integral 
over all spatial locations and all interactions
\begin{eqnarray}
 {\mathcal F}&=&\int_\Omega dx\,W\left[V(x,t]\right)\nonumber\\
&&-\frac{1}{2}\int_\Omega dx\, \int_\Omega dy\,K(x-y)V(x,t)V(y,t)\label{eqn_F}\\
 &=&\int_\Omega dx\,W\left[V(x,t)\right]-\frac{1}{2} V(x,t)\sum_{n=0}^\infty(-1)^nK_n\frac{\partial^n V(x)}{\partial x^n}\nonumber\\
 &=&\int_\Omega dx\,W\left[V(x,t)\right]-\frac{1}{2} V(x,t)\nonumber\\
&&\times\left(K_0+K_2\frac{\partial^2 V(x)}{\partial x^2}+K_4\frac{\partial^4 V(x)}{\partial x^4}+\cdots\right)\nonumber
\end{eqnarray}
with $-\delta W/\delta V=h[V]$ and $\Omega$ denotes the real axis. In the third line, we applied
the identity (\ref{eqn_PDEexp}). Then the specific choices applied in the previous paragraph
yields the well-known energy functional of the SHE~\cite{Swift+Hohenberg,Swift+Hohenberg2,Cross80}
\begin{eqnarray}
 {\mathcal F}&=&\int_\Omega dx\,\frac{1}{4}V^4(x,t)\nonumber\\
&&+\frac{1}{2}\int_\Omega dx \,V(x,t)\left(-\varepsilon+\left(1+\frac{\partial^2}{\partial x^2}\right)^2\right)V(x,t).\nonumber\\
&&\label{eqn_SHfunc}
\end{eqnarray}
Corresponding to this derivation of the SHE from general physical considerations, 
we conclude that the SHE represents a specific description of a general spatial system. This 
view angle on the equation is supported by its various motivations in physics mentioned above.\\
In addition to the previous discussion, Eq.~(\ref{eqn_F}) is the Lyapunov functional of Eq.~(\ref{eqn_SH_nonlocal}) as
\begin{eqnarray*}
\frac{d}{dt}{\mathcal F}=-\int_\Omega dx\,\left(\frac{\partial V(x,t)}{\partial t}\right)^2<0~.
\end{eqnarray*}
Then the question arises whether there exists an energy functional 
in the case of finite propagation speeds. Corresponding to the previous treatment, the first ansatz 
for a possible energy functional would contain the terms $K(z)V(x,t)V(x-z,t-\tau(z))$ with 
the distance $z=x-y$ between two spatial locations and the propagation delay $\tau(z)\ne 0$ taken 
from Eq.~(\ref{eqn_final}). Then applying the identity (\ref{eqn_PDEexp}) in the time domain, 
we obtain
\begin{eqnarray*}
&&K(z)V(x,t)V(x-z,t-\tau(z))=\nonumber\\
&&K(z)V(x,t)\sum_{n=0}^\infty \frac{1}{n!}\left(-\frac{|z|}{c}\right)^n\frac{\partial^n V(x-z,t)}{\partial t^n}~.
\end{eqnarray*}
We observe that the energy of a pair of elements at distance $z$, whose interaction is delayed by the 
finite propagation speed $c$, depends on the temporal derivatives of the elements.
Subsequently the energy functional of the system ${\mathcal F[V,\partial V/\partial t,\ldots]}$ 
would depend on the temporal derivatives of the field and the generalized Ginzburg-Landau equation 
(\ref{eqn_GL}) is not applicable. Preliminary results on the different derivation of
 Eq.~(\ref{eqn_SH_nonlocal}) including propagation delays by another variational principle seems
promising. However, the discussion of these results would exceed the major aim of 
the present work and we refer the reader to forthcoming work.

\subsection{Kuramoto-Sivashinsky equation}\label{subsec_KS}
At last we focus on the Kuramoto-Sivashinsky equation~\cite{Hyman86,Cross93} which allows for 
the study of various phenomena in fluids and solids as e.g. the phase turbulence in
 fluids~\cite{Shraiman86}, the thermal diffusive instabilities of flame fronts~\cite{Frisch_etal86},
the directional solidification in alloys~\cite{Novick-Cohen87} or the interface instability during 
the application of industrial beam cutting techniques~\cite{Friedrich_PRL00}. Here, the terms in
 Eq.~(\ref{eqn_final}) are chosen as $\hat{T}=\partial/\partial t$, $g(V)=g_1V,~f(V)=f_2V^2$ with $g_1,~f_2<0$. Now the kernel $K(x)$ is  
non-symmetric and short-ranged, i.e. $K_n\to 0,~n>1$ while the kernel $L(x)$ is chosen symmetric. Hence 
the corresponding kernel moments obey $L_{2n+1}=0$ and $L(x)$ describes 
long-range interaction with $L_n\to 0, n>4$. Then the application of Eq.~(\ref{eqn_PDEexp}) leads to
\begin{eqnarray*}
\frac{\partial V(x,t)}{\partial t}&=&(L_0g_1-h_0)V(x,t)+L_2g_1\frac{\partial^2}{\partial x^2}V(x,t)\nonumber\\
&&+L_4g_1\frac{\partial^4}{\partial x^4}V(x,t)+2K_1f_2V\frac{\partial V}{\partial x}
\end{eqnarray*}
while $h(V)$ has been chosen as $h(V)=-K_0f_2V^2-h_0V,~h_0\ge 0$. Finally the re-scaling $t\to (L_4/L_2^2|g_1|)t$
and $x\to\sqrt{L_4/L_2}x$ yields 
\begin{eqnarray}
\frac{\partial V(x,t)}{\partial t}&=&-\eta V-\frac{\partial^2}{\partial x^2}V(x,t)-\frac{\partial^4}{\partial x^4}V(x,t)-V\frac{\partial V}{\partial x}\nonumber\\
\label{eqn_KS}
\end{eqnarray}
with $\eta=|g_1|L_0+h_0$ and the additional condition $f_2=L_2\sqrt{L_2/L_4}/2K_1|g_1|$.
Equation~(\ref{eqn_KS}) is called generalized Kuramoto-Sivashinsky equation~\cite{Cross93} and $\eta=0$ 
represents the original Kuramoto-Sivashinsky equation. Hence systems obeying the Kuramoto-Sivashinsky 
equation involve symmetric spatial interactions up to order $O(\sigma^5)$ and non-symmetric spatial 
interactions of order $O(\sigma)$. It is interesting to note that this model includes both non-symmetric and symmetric spatial interactions while the reaction-diffusion equation and the Swift-Hohenberg equation involve symmetric spatial interactions only.

\subsection{Validity study}\label{subsec_validity}
After the previous examinations, the question arises which specific kernel functions may fulfill the appropriate conditions 
in the three models. To this end, we apply the Fourier transform to the nonlinear function $S(V(x))$ in (\ref{eqn_PDEexp}) and obtain the spatial mode expansion 
\begin{eqnarray}
\int_{-\infty}^{\infty}dyK(x-y)S(V(y,t))=\int_{-\infty}^\infty dk\tilde{S}(k,t)e^{ikx}\sum_{n=0}^\infty f_n(k)\nonumber\\
\label{eqn_SFourier}
\end{eqnarray}
with the sequence $f_n(k)=(-ik)^nK_n$. In the following, we abbreviate $f_n=f_n(k)$ for 
convenience. As we will see later, $f_n(k)$ represents the contribution of interactions at order $n$ to the spatial modes of the system.\\
To obtain convergence of the series expansion in (\ref{eqn_PDEexp}), the ratio of subsequent sequence terms 
$|f_n|$ has to be smaller than unity. This means this condition reads $|f_{n+2}|/|f_n|<1$ for $n$ even 
and $|f_{n+1}|/|f_n|<1$ for all $n$ for symmetric and non-symmetric kernels, respectively. \\
To be more specific let us focus on a specific symmetric kernel function. Since the kernel represents the
 probability density function of the spatial interactions between two locations, the central limit theorem 
supports the choice of the Gauss function $K(x)=\exp(-x^2/2\sigma^2)/\sqrt{2\pi}\sigma$ with the spatial range
constant $\sigma$. Then the previous convergence condition leads to $\sigma^2<(n+2)/k^2, n=2,4,6,\ldots$~. 
If there is a maximum spatial
frequency $|k_m|$, i.e. the Fourier transform $\tilde{S}(|k|)$ is negligible for $|k|>|k_m|$, 
then the spatial interaction range $\sigma$ is limited from above by $\sqrt{n+2}/|k_m|$~. Subsequently, 
the spatial interaction range $\sigma$ is delimited to the interval $0<\sigma^2<1/k_m^2$ which guarantees 
the Taylor expansion convergence in (\ref{eqn_PDEexp}). We point out that the limitation to spatial modes is 
present in various physical systems. For instance, spatial systems near a phase transition exhibit a hierarchy 
of time scales of spatial modes with corresponding prominent spatial frequencies. In this 
case the time evolution of the Fourier transforms $\tilde{S}(k,t)$ at spatial frequencies far from the prominent ones are negligible and the prominent frequencies define the maximum spatial frequency $|k_m|$. However,  
if there is no maximum spatial frequency, i.e. $|k_m|\to\infty$, it is $\sigma\to 0$ and the convergence  
condition yields $K(x)\to\delta(x)$. Hence for an unlimited range of spatial frequencies, the expansion
 (\ref{eqn_PDEexp}) is valid for local interactions only and $K_n\to 0$ for $n>0$. \\
Now let us interpret the latter results. It turns out that the convergence condition for 
Gaussian kernels leads to the dependence of the maximum spatial range on the interaction order $n$. 
For instance if the expansion in Eq.~(\ref{eqn_PDEexp}) approximates local interactions but no higher orders of nonlocality (cf. the previous discussion of the reaction-diffusion equation), it is 
$|K_{n+2}/K_n|\ll 1$ for $n=2,4,\ldots$ and, subsequently, $\sigma^2<(2+2)/k^2=4/k^2$. Hence, 
a maximum spatial frequency $|k_m|$ yields the validity condition $0<\sigma^2<4/k_m^2$. In case of spatial
interactions at higher orders of nonlocality (cf. the previous discussion of the Swift-Hohenberg and the 
Kuramoto-Sivashinsky equation), it is 
$|K_{n+2}/K_n|\ll 1$ for $n=4,6,\ldots$. This means that the validity condition now reads  
$0<\sigma^2<(4+2)/k_m^2=6/k_m^2$. Thus the spatial interaction range $\sigma$ may be larger 
than in the diffusion model. \\
Since the Kuramoto-Sivashinsky equation involves non-symmetric interactions, let us also examine briefly the
 kernel function $K(x)=\delta(x-x_0)$ with the non-symmetric shift $x_0\ne 0$. We find $K_n=x_0^n/n!$ and the
 convergence condition reads $x_0<(n+1)/|k_m|$. Subsequently, the Kuramoto-Sivashinsky equation is an
 approximation of (\ref{eqn_final}) if $x_0<2/|k_m|$ and if a maximum spatial frequency exists.

\section{Linear analysis}\label{sec_stability}
Now let us examine the stationary state and its linear behavior for a finite propagation speed $c$.
In case of the stationary state $V(x,t)=V_0$, Eq.~(\ref{eqn_final}) recasts to 
\begin{eqnarray}
T(0)V_0&=&h(V_0)+\kappa_K f(V_0)+\kappa_L g(V_0) ,\label{eq_V0}\\
~\kappa_K&=&\int_{-\infty}^\infty K(x)dx,~\kappa_L=\int_{-\infty}^\infty L(x)dx.\nonumber
\end{eqnarray}
with $\kappa_K=\int K(x)dx,~\kappa_L=\int L(x)dx$. Considering linear deviations 
$u(x,t)=V(x,t)-V_0\sim e^{\lambda t+ikx}$, Eq.~(\ref{eqn_final}) reads
\begin{eqnarray}
T(\lambda)=s_h+\int_{-\infty}^{\infty}dz M(z)e^{-\lambda|z|/c}e^{-ikz}\label{eqn_Llambda}
\end{eqnarray}
with $M(z)=s_fK(z)+s_gL(z)$ and $s_h=\delta h/\delta V,~s_f=\delta f/\delta V,~s_g=\delta g/\delta V$ computed at $V=V_0$, while $\delta/\delta V$ denotes the functional derivative.
In the following, the propagation delay $c$ is assumed being large but finite and the approximation $\exp(-\lambda|z|/c)\approx 1-\lambda|z|/c+\lambda^2(|z|/c)^2$ holds. For the characteristic spatial scale of the system
 $\sigma$, we introduce the characteristic propagation delay $\tau=\sigma/c$ which represents the delay time 
caused by the finite propagation speed. Thus the previous approximation holds for small propagation delays $\tau\ll 1/\lambda$.  Subsequently 
expanding the exponential $\exp(-ikz)$ into a power series we obtain 
\begin{eqnarray}
&&T(\lambda)+\frac{\lambda}{c}\sum_{n=0}^\infty(-ik)^nP_n-\frac{\lambda^2}{c^2}\sum_{n=0}^\infty(-ik)^nQ_n\nonumber\\
&&=s_h+\sum_{n=0}^\infty(-ik)^nM_n\label{eqn_Tlambda-final}
\end{eqnarray}
with the kernel moments $M_n$ defined previously and $P_n=\int_\Omega dz M(z)|z|z^n/n!,~Q_n=\int_\Omega dz M(z)|z|^2z^n/n!$. Preliminary computations of $M_n, P_n$ and $Q_n$ for Gaussian kernels yield 
$M_n\sim O(\sigma^n)$, $P_n\sim O(\sigma^{n+1})$ and $Q_n\sim O(\sigma^{n+2})$ \\
Equation~(\ref{eqn_Tlambda-final}) defines the stability condition for general spatial interactions and short
 propagation delays. Additional more detailed studies show that large propagation speeds $c$ with 
$1/c^2\approx 0$ and vanishing odd kernel moments $M_{2n+1}=0,~P_{2n+1}=,~Q_{2n+1}=0$ lead to real Lyapunov
exponents and thus may yield stationary bifurcations. In contrast oscillatory bifurcations may emerge for non-vanishing odd moments
 only. By recalling the definitions of these moments, we find that symmetric and non-symmetric spatial
 interactions may yield stationary and non-stationary bifurcations, respectively.
In case of smaller propagation speeds $1/c^2\ne 0$, this classification does not hold anymore and
oscillatory bifurcations may occur for both symmetric and non-symmetric kernels.

\subsection{The reaction-diffusion equation}
To illustrate the latter results, first let us study a specific reaction-diffusion equation, namely the homogeneous diffusion equation also known as
Ficks Second Law. According to the previous discussion, Eq.~(\ref{eqn_RD}) represents the diffusion equation with $\bar{h}=0$ yielding 
$h(V)=-f_1V$. Further Eq.~(\ref{eq_V0}) yields the stationary solution $V_0=0$ and the parameters $s_f=f_1,~s_h=-f_1$, while the spatial interactions are given by $K(z)\ne 0$ and $L(z)=0$, i.e. 
$M(z)=f_1K(z)$. \\
According to the previous discussion of (\ref{eqn_SFourier}), Eq.~(\ref{eqn_PDEexp}) represents a spatial mode
 expansion in orders of the characteristic spatial scale $\sigma$. That 
is, in the case of the diffusion equation, the order expansion is truncated at $O(\sigma^3)$ yielding $M_0=f_1\sim O(1),~M_2=f_1\sigma^2/2\sim O(\sigma^2)$
 while all odd and higher kernel moments $M_{2n}$ vanish with $n\ge 2$. Further we find 
 $P_0=f_1\sqrt{2}\sigma/\sqrt{\pi},~P_1=0,~P_2=f_1\sqrt{2}\sigma^3/\sqrt{\pi}$
 with $P_{n}\approx 0~\forall~n>3$ and $Q_0=f_1\sigma^2,~Q_n\approx 0~\forall~n>0$. Here it is $f_1=2D\sigma^2$ 
 and $D$ represents the diffusion constant. Subsequently in the case of $c\to\infty$ 
  Eq.~(\ref{eqn_Tlambda-final}) yields  $\lambda(k)\approx -Dk^2$ which is the well known dispersion 
relation of the diffusion equation. 
\begin{figure}
\includegraphics[width=8cm]{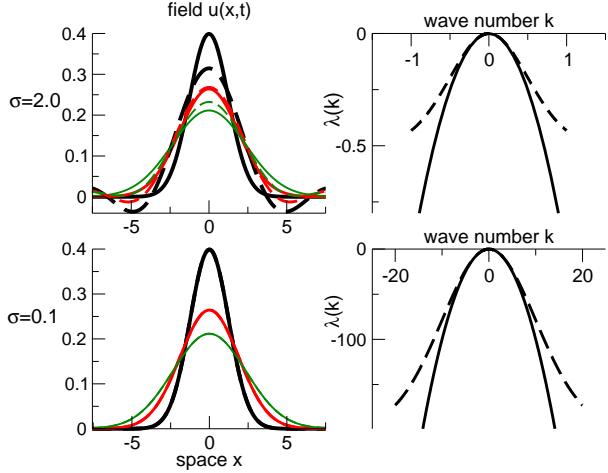}
\caption{(Color online)~The field of the diffusion equation and the Lyapunov exponent for different spatial ranges $\sigma$. The dashed lines 
represent the results from the diffusion PDE, while the solid lines denotes the results from the IDE model. In case
of $\sigma=0.1$, the reconstructed field $u(x,t)$ obtained from the IDE can not be distinguished from the field of 
the original diffusion equation. In addition the colors
in the left panel denote subsequent times $t=0$~(black, very thick line),~$t=3$~(red, thick line) and $t=6$~(green, thin line). That is the field 
$u(x,t)$ flattens at increasing time. \label{fig_lambda}}
\end{figure}
Figure~\ref{fig_lambda} compares $\lambda$ and $\lambda_{RD}$ and the corresponding resulting field 
activity for two values of $\sigma$. We observe that the IDE model is a good approximation to the diffusion
equation for $\sigma=0.1$, while the the approximation is much worse for $\sigma=2$. This finding 
confirms our previous results on the convergence condition. Here the maximum value of $|k|$ is $2/\sigma$
according to the convergence condition $\sigma<2/k$ derived in the previous paragraphs.

\begin{figure}
\includegraphics[width=8cm]{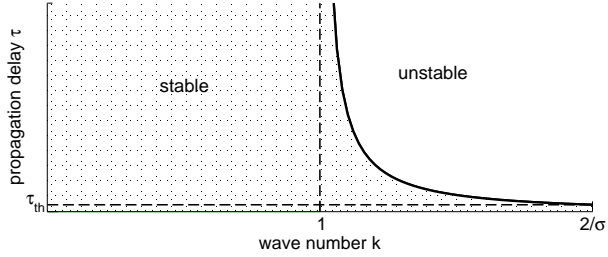}
\caption{Illustrated stability diagram of the extended diffusion equation. The vertical axis $k=1$ represents
the asymptote of the instability regime.\label{fig_RD_tauk}}
\end{figure}

Now let us turn to finite propagation speeds $c<\infty$. Considering the corresponding orders of 
$M_n,~P_n$ and $Q_n$, Eq.~(\ref{eqn_Tlambda-final}) reads
\begin{eqnarray}
\lambda+\frac{\lambda}{c}(P_0-P_2k^2)-\frac{\lambda^2}{c^2}Q_0=s_h+M_0-M_2k^2.\label{eqn_RDlambda1}
\end{eqnarray}
For small propagation delays $(\lambda\tau)^2\to 0$ we find the Lyapunov exponent
\begin{eqnarray}
\lambda_{RD}(k)=\frac{-Dk^2}{1+\alpha\tau(1-\sigma^2k^2)}\label{eqn_RDlambda}
\end{eqnarray} 
with $\alpha=f_1\sqrt{2/\pi}$. By virtue of the wave number limit $|k|<2/\sigma$, there is a critical propagation 
delay $\tau_{th}=1/3\alpha$ and a corresponding critical propagation speed $c_{th}=3\alpha\sigma$. Then 
small delays $\tau<\tau_{th}$ retain the field stability, while large delays $\tau>\tau_{th}$ yield 
linear unstable modes $k$ with $\lambda_{RD}(k)>0$. Figure~\ref{fig_RD_tauk} shows the corresponding 
stability diagram of the system.\\
Moreover, Eq.~(\ref{eqn_RDlambda}) may be interpreted as if it originates from the modified diffusion equation
\begin{eqnarray}
\frac{\partial u(x,t)}{\partial t}+\frac{\alpha\sigma}{c}\left(1+\sigma^2\frac{\partial^2}{\partial x^2}\right)\frac{\partial u(x,t)}{\partial t}=D\frac{\partial^2 u(x,t)}{\partial x^2}\label{eqn_newmodel}
\end{eqnarray}
This equation is an extension of the standrad diffusion equations considering finite propagation speeds.
It applies in media whose propagation delay on the typical length scale of the system is not negligible and, for instance, which 
show delayed spread of space-time activity. We mention results from fast hot  
 pulses in plasma~\cite{Lazzaro+Wilhelmsson98}, whose investigation
 indicates nonlocal effects in the reaction-diffusion mechanism. Moreover,
 there is 
the debate on the non-Fourier heat conduction in nonhomogeneous materials 
which show such delayed temporal activity at measurement points~\cite{Mitra_etal95,Tzou+Chen98,Herwig+Beckert00}.
\begin{figure}[b]
\includegraphics[width=8cm]{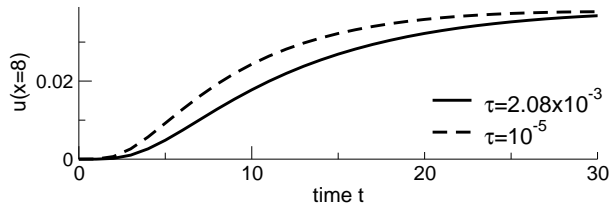}
\caption{Temporal activity of the modified diffusion equation at spatial location $x=8$ for two different propagation delays, i.e. propagation speeds. The solutions are computed analytically with initial condition $u(k,t=0)=\exp(-k^2/2\cdot0.8^2)/\sqrt{2\pi 0.8^2}$~.\label{fig_RDfield-t}}
\end{figure}
In order to compare our model to the results found in experiments~\cite{Mitra_etal95},
 Fig.~\ref{fig_RDfield-t} shows the temporal activity at a spatial point. We observe a 
prominent delay of the activity at large delays,
 while $\tau\to 0$ yields the Fourier case. This finding is irrespective of the spatial location used and
 coincides qualitatively to experimental findings. Moreover, the derived model (\ref{eqn_newmodel}) contrasts 
to the Cattaneo equation~\cite{Cattaneo58,Metzler+Nonnenmacher98}, which represents a telegraphic equation and extends
 the parabolic diffusion equation to a hyperbolic partial differential equation by an additional second 
temporal derivative and thus involves finite propagation speed.
Originally the Cattaneo extension has been proposed according to heuristic arguments while our model is derived
 from the explicit treatment of propagation delay. \\
At last, let us briefly focus on the case of larger propagation delays $(\lambda\tau)^3\to 0$ yielding the
implicit equation $-\lambda^2\tau^22D/\sigma^2+\lambda(1+\alpha D\tau(1-\sigma^2k^2))-Dk^2=0$. Simple 
calculus reveals $Re(\lambda)>0$ for some $k$ and larger delays destabilize the field similar to the previous
case of small delays.

\subsection{The Swift-Hohenberg equation}
Now let us turn to the Swift-Hohenberg model which involves long-range spatial interactions.
It has been studied as a generic model for spatial pattern formation in 
various physical systems~\cite{Bestehorn+Haken90,Lega94,Brazovskii75}. By virtue of its broad 
applicability, this paragraph aims to extend it by finite propagation delay effects. 
The stationary state is given by Eq.~(\ref{eq_V0}) and we find one or three solutions. The single  
solution $V_0=0$ leads to the parameters $s_f=-|f_1|$ and $s_h=a$. Further the spatial 
interactions are given by $K(z)\ne 0$ and $L(z)=0$ yielding $M(z)=-|f_1|K(z)$. \\
According to the discussions in section~\ref{subsec_validity}, Eq.~(\ref{eqn_PDEexp}) breaks 
off at order $O(\sigma^5)$ and we obtain the expansion terms
\begin{eqnarray*}
&&M_0=-|f_1|,~M_2=-|f_1|\sigma^2/2,~M_4=-|f_1|\sigma^4/8\\
&&P_0=-|f_1|\sqrt{\frac{2}{\pi}}\sigma,~P_2=-|f_1|\sqrt{\frac{2}{\pi}}\sigma^3,~P_4=-|f_1|\frac{\sqrt{2}}{3\sqrt{\pi}}\sigma^5\\
&&Q_0=-|f_1|\sigma^2~,Q_2=-|f_1|\frac{3}{2}\sigma^4
\end{eqnarray*}
while all odd and higher terms $M_n,~P_n,~Q_n$ vanish. Subsequently Eq.~(\ref{eqn_Tlambda-final}) yields 
for $c\to\infty$ 
\begin{eqnarray*}
\lambda_\infty=\frac{|f_1|}{2}\left(\epsilon-1+2l^2-l^4\right)\label{eqn_SH_charac}
\end{eqnarray*}  
with the scaled wave number $l=\sigma k/\sqrt{2}$ and $\varepsilon$ taken from 
section~\ref{subsec_SH}. Here $\lambda_\infty$ represents the Lyapunov exponent 
of the linearized original Swift-Hohenberg equation (\ref{eqn_SH}) and thus defines the 
linear stability of the underlying spatial system. To be more detailed, for 
$\varepsilon<0$ the system is linear stable with $\lambda_\infty<0$ for all $l$, while
 $\varepsilon>0$ yields $\lambda_\infty>0$ for $1-\sqrt{\varepsilon}<l<1+\sqrt{\varepsilon}$. In other words, 
 a Turing instability emerges for $\varepsilon>0$ at scaled wave numbers $|l|\approx 1$.\\
Now let us turn to the case $c<\infty$. For small propagation delays $(\lambda\tau)^2\to 0$ we find
\begin{eqnarray}
&&\lambda_{SH}=\frac{\lambda_\infty}{1-\alpha(1-2l^2+4l^4/3)\tau}\label{eqn_SH_lambda}
\end{eqnarray}  
with $\alpha=|f_1|\sqrt{2/\pi}$. Here $\lambda_{SH}$ represents the 
Lyapunov exponent of the Swift-Hohenberg equation subject to large propagation speeds.
It is important to mention that the validity criteria $0<\sigma^2<6/k_m^2$ derived in
 section~\ref{subsec_validity} delimits the scaled wave number to $-\sqrt{3}<l<\sqrt{3}$. 
Taking into account this constraint, closer examinations of Eq.~(\ref{eqn_SH_lambda}) reveal 
the critical propagation delay $\tau_{th}=1/7\alpha$ and the corresponding critical propagation 
speed $c_{th}=7\alpha \sigma$. For $\tau<\tau_{th}$, the denominator of
$\lambda_{SH}$ is positive for all valid wave numbers and thus $\lambda_\infty$ defines the
 stability of the system. Hence the propagation delay does not affect the stability 
of the system.\\
 In contrast large propagation delays $\tau>\tau_{th}$ yield a 
 negative denominator for a half-band of wave numbers $|l|>l_{th}$ and thus change the sign of 
the Lyapunov exponent. Here the critical 
 wave number $l_{th}$ represents the root of the denominator in Eq.~(\ref{eqn_SH_lambda}). 
\begin{figure}
\includegraphics[width=8cm]{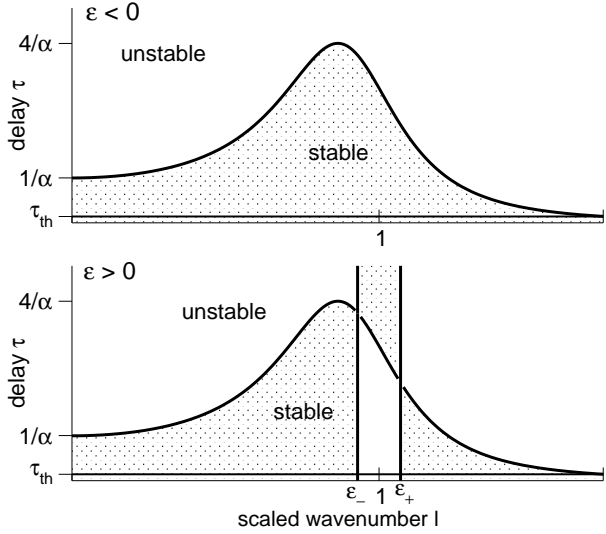}
\caption{Illustrated stability diagram of the extended Swift-Hohenberg equation involving propagation delays. For 
$\varepsilon<0$ (top panel) the underlying system is stable for $\tau<1/\alpha(1-2l^2+4l^4/3)$ and unstable
otherwise. For $\varepsilon>0$ (bottom panel) the stability is the same as for $\varepsilon<0$ besides the wave number band
$\varepsilon_-<1<\varepsilon_+,~\varepsilon_\pm=\pm\sqrt{\varepsilon}$. For short delays 
$\tau<1/\alpha(1-2l^2+4l^4/3)$ the underlying system is linear unstable in this wave number band, while
the modes in this band are stable otherwise. In both latter cases $\tau<\tau_{th}$ guarantees the linear stability of the underlying system.~\label{fig_SH_taul}}
\end{figure}
Figure~\ref{fig_SH_taul} shows the stability diagram of Eq.~(\ref{eqn_SH_lambda}). We observe the unstable
 half-band for $\tau>\tau_{th}$ for all $\varepsilon$, 
  while $\tau>1/\alpha>\tau_{th}$ yields an additional unstable band of wave numbers around 
 $l=0$. For $\varepsilon<0$ the stable band of spatial modes vanish for large propagation delays
 $\tau>4/\alpha$ and the system is totally destabilized, while for $\varepsilon>0$ increased propagation 
delays stabilize the band of unstable Turing-modes. This stabilization by propagation delays represents 
a novel effect  
near Turing instabilities. In addition recall that the previous 
treatment around $l_{th}$ reflects $|\lambda|\to\infty$ and thus conflicts with the previous 
assumption $(\lambda\tau)^2\approx 0$. Subsequentially higher polynomial orders in $\lambda$ are 
necessary for closer examinations. \\
Finally Eq.~(\ref{eqn_SH_lambda}) may be interpreted as if it originates from the extended 
Swift-Hohenberg equation
\begin{eqnarray}
&&\frac{\partial V(x,t)}{\partial t}-\frac{\alpha\sigma}{c}\left(1+2\frac{\partial^2}{\partial x^2}+\frac{4}{3}\frac{\partial^4}{\partial x^4}\right)\frac{\partial V(x,t)}{\partial t}\nonumber\\
&&=\epsilon V(x,t)-V^3(x,t)-\left(1+\frac{\partial^2}{\partial x^2}\right)^2V(x,t)~\label{eqn_SHext}
\end{eqnarray}
with the re-scaled time $t\to 2t/|f_1|$. This equation represents an 
extension of the well-knwon Swift-Hohenberg equation involving a finite propagation speed. In the case of 
a infinite propagation speed $c\to \infty$ the additional second term on the left hand side 
of Eq.~(\ref{eqn_SHext}) vanishes and the original model equation is re-gained.
Further the finite propagation speed leads to the coupling term between spatial and 
temporal derivatives in Eq.~(\ref{eqn_SHext}) reflecting the space-dependent delay in Eq.~(\ref{eqn_final}).

\subsection{The Kuramoto-Sivashinsky equation}
\begin{figure}
\includegraphics[width=7cm]{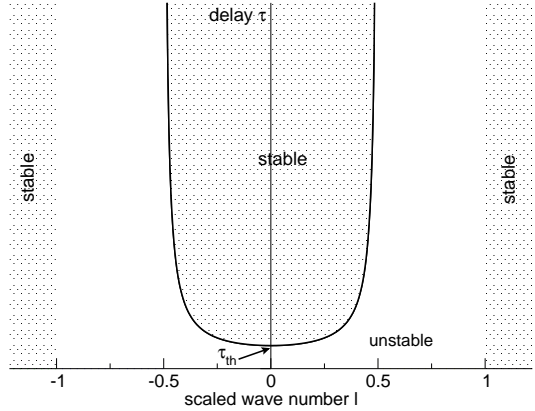}
\caption{The illustrated stability diagram of the extended Kuramoto-Sivashinsky equation subject. The center
stability region does not exceed the interval $[-0.5,0.5]$ on the horizontal axis.~\label{fig_KStaul}}
\end{figure}

Finally let us investigate the Kuramoto-Sivashinsky equation by following the same steps as in the previous
examples. Now Eq.~(\ref{eqn_Tlambda-final}) reads for small propagation delays 
 $(\lambda\tau)^2\approx 0$
\begin{eqnarray*}
&&\lambda+\frac{\lambda}{c}(P_0-k^2P_2)=s_h+M_0-ikM_1-k^2M_2+k^4Q_2.
\end{eqnarray*}
with 
$M_0=s_f+s_g,~M_1=s_fx_0,~M_2=s_g\sigma^2/2,~M_4=s_g\sigma^4/8,~P_0=s_fx_0+s_g\sqrt{2}\sigma/\sqrt{\pi},~P_2=s_g\sqrt{2}\sigma^3/\sqrt{\pi}$ and $M_n,~P_m\to 0$ for $n>4,~m>2$.
The stationary solution is taken from (\ref{eq_V0}) and reads $V_0=0$ yielding $s_f=0,~s_g=g_1$ and 
the Lyapunov exponent 
\begin{eqnarray}
\lambda_{KS}=\frac{-\eta+l^2-l^4}{1-\tau\alpha(1-4l^2)}\label{eqn_KSext}
\end{eqnarray} 
with $\alpha=|g_1|\sqrt{2/\pi}$,~$l=\sigma k/2$ and $\lambda_{KS}=\lambda/2|g_1|$. We point out that the
 re-scalings $k\to l$ and $\lambda\to\lambda_{KS}$ are identical to the re-scalings of space and time in 
section~\ref{subsec_KS}.\\
\begin{figure}[b]
\includegraphics[width=7cm]{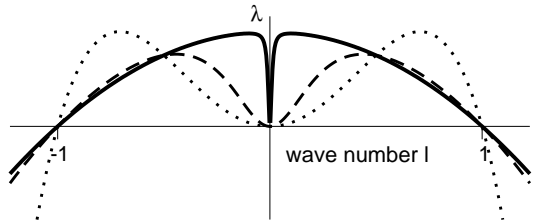}
\caption{The Lyapunov exponent of the Kuramoto-Sivashinsky equation subject to the scaled wave number $l$. The different line styles encode the different propagation delays $\tau=0$~(dotted line), 
$\tau=2.0$~(dashed line) and $\tau=2.499$ (solid line). It is $\tau_{th}=2.5$.~\label{fig_KSlambda}}
\end{figure}
Now let us examine (\ref{eqn_KSext}) in some more detail. For a vanishing propagation delay $\tau=0$
 $\lambda_{KS}$ becomes the Lyapunov exponent well-known for the original Kuramoto-Sivashinsky model. This 
means the stationary solution is linear unstable at wave numbers $-1<l<1$ while it is linear stable otherwise. 
In case of $\tau\ne 0$, the propagation delay changes the Lyapunov exponent and may destabilize the system. 
It turns out that there is a critical propagation delay $\tau_{th}=1/\alpha$ and a corresponding a critical
propagation speed $c_{th}=\alpha\sigma$. For small propagation delays $\tau<\tau_{th}$, the Lyapunov 
exponents are modified but the system stability is retained, see Fig.~\ref{fig_KSlambda}). In contrast 
large propagation delays $\tau>\tau_{th}$ lead to a sign inversion of the Lyapunov exponent for
 $-l_c<l<l_c$ with $l_c^2=1-1/\alpha\tau$. Figure~\ref{fig_KStaul} summarizes the stability conditions 
in a diagram. However $\tau\to\tau_{th}$ yields $\lambda\to\infty$ for $l\to l_c$ and thus does not fulfill the previous condition $(\lambda\tau)^2\approx 0$.\\
Finally, Eq.~(\ref{eqn_KSext}) may be interpreted as if it originates from the PDE
\begin{eqnarray}
&&\frac{\partial V(x,t)}{\partial t}-\frac{\sigma\alpha}{c}\left(1+4\frac{\partial^2}{\partial x^2}\right)\frac{\partial V(x,t)}{\partial t}=\nonumber\\
&&-\eta V-\frac{\partial^2}{\partial x^2}V(x,t)-\frac{\partial^4}{\partial x^4}V(x,t)-V\frac{\partial V}{\partial x}
\end{eqnarray}
This equation represents an extension of the Kuramoto-Sivashinsky equation now involving large propagation
speeds $c$. We observe that $c\to\infty$ yields the original Kuramoto-Sivashinsky equation, while the 
finite propagation speed leads to the additional second term on the left hand side. 

\section{Conclusion}
The present work has introduced a nonlocal integral-differential equation model 
which generalizes three important partial differential equations. Moreover we have 
shown how to consider large propagation speeds in these models and have quantified  
the propagation delay as the fraction of spatial interaction range and propagation speed. 
The subsequent linear stability analysis turns out to be dependent on this quantity 
and reveals critical propagation delays and corresponding propagation speeds. We 
find that large propagation speeds slow down the activity spread in diffusion systems 
and thus allows for the explanation of non-Fourier behavior in non-homogeneous systems.
Further finite propagation speeds may destabilize the dynamics of Swift-Hohenberg equation
in the stable regime of the original equation, while it may stabilizes an occurring Turing 
instability. Finally finite propagation speeds may stabilize the dynamics of the 
Kuramoto-Sivashinsky equation. For all three models novel partial differential equations 
have been formulated which incorporate the propagation delay effects. We are convinced that 
propagation delays play an important role in media showing ultra-fast phenomena and that 
the obtained results shall allow for their modeling.

\section{Acknowledgment}
The author thanks the anonymous referee for valuable comments and acknowledges the financial support from the Deutsche Forschungsgemeinschaft 
(grant Sfb-555).

\end{document}